\documentclass[12pt]{article}
\usepackage{geometry}
\usepackage{amsmath,amssymb,amsfonts,amsthm,graphicx,epstopdf}
\usepackage{url,setspace,nicefrac,bbm, natbib}
\DeclareGraphicsExtensions{.eps,.pdf}

\allowdisplaybreaks

\newcommand{\var}{\text{Var}}
\newcommand{\cov}{\text{Cov}}

\newcommand{\ev}{\ensuremath{\mathbbm{E}}}

\author{Amy Willis}
\title{Species richness estimation with high diversity but spurious singletons}
\date{}

\begin{document}
\maketitle

\section*{Informal note from the author}
The method described in this paper has been available via my \url{R} package \url{breakaway} since August 2014. I am not pursuing publication of this note, but due to the amount of interest that I have received since its release I felt the microbial community deserved an explanation of the method. The punchline is that species richness estimation without a reliable singleton count is essentially impossible. I mean this in the sense that the sampling variability of the problem explodes (there is almost infinite variance), even under parsimonious models. This method is interesting for demonstrating this point but I would generally not recommended its use (nor the use of any method for $\alpha$-diversity estimation without singletons!) for drawing conclusions about population diversity. Please remember that quoting a large standard error is an honest statement about sampling variability, and that failing to quote any measure of variability is sloppy science (at best). As with all of my methods and software, please feel free to contact me with questions, extension requests and enquiries at \url{adw96@cornell.edu}; it is my great pleasure to interact with microbial ecologists and facilitate improved information exchange between statisticians and scientists. 

\section*{Abstract}
The presence of uncommon taxa in high-throughput sequenced ecological samples pose challenges to the microbial ecologist, bioinformatician and statistician. It is rarely certain whether these taxa are truly present in the sample or the result of sequencing errors. Unfortunately, $\alpha$-diversity quantification relies on accurate frequency counts, which can rarely be guaranteed. We present a species richness estimation tool which predicts both the number of unobserved taxa and the number of true singletons based on the non-singleton frequency counts. This method can be treated as either inferential (for formally estimating richness) or exploratory (for assessing robustness of the richness estimate to the singleton count). If the estimate, called \url{breakaway_nof1}, is comparable to other richness estimators, this provides evidence that the richness estimate is robust to the level of quality control (eg. chimera-checking) employed in pre-processing. The function \url{breakaway_nof1} is freely available from CRAN via the R package breakaway.

\section*{Introduction}

Next-generation sequencing has greatly increased our understanding of microbial community structure and function. The Human Microbiome Project is illuminating the microscopic inhabitants of our bodies, while the Earth Microbiome Project is revealing the incredible diversity of microorganisms in the unlikeliest of landscapes. It is now common for a single study to identify millions of distinct operational taxonomic unit (OTU) clusters, especially in highly heterogeneous environments such as soil and water. 

The size of these datasets complicates both their processing and their analysis. Amplification and sequencing of genetic markers introduces genetic noise, and the bioinformatics community continues debate over the best tools for preprocessing compositional data. Of particular controversy is the appropriate abundance of ``singletons'' in a dataset. Global singletons (OTU clusters that only appear once even across replicates) may be false: introduced in the sequencing process rather than reflective of the environment under study. Local singletons (individuals which appear only once per sample but in multiple replicates) are more controversial to discard. 

The high dimensionality of community composition data demands appropriate summary statistics, and $\alpha$-diversity (species richness and evenness) is commonly studied. Unfortunately, estimation of $\alpha$-diversity is rarely statistically rigorous. For example, quoting standard errors on evenness index estimates is uncommonly rare. Furthermore, measurement error in the number of singletons exerts high influence on $\alpha$-diversity metrics. 

\cite{first} discuss species richness estimation with measurement error, however no software exists that implements the proposed solutions. Here, we present a new method as well as a freely available implementation. Based on the structure of the non-singleton abundances, our method predicts the number of singleton OTUs and the number of unobserved OTUs in the population under study. Standard errors and fit diagnostics are also given. This method provides a first statistically rigorous richness software implementation that accounts for measurement error induced in the bioinformatics pipeline.

\section*{Theory \& Method}

Our approach to predicting the unobserved richness as well as the true singleton count extends recent work on modelling frequency counts via their ratios by \cite{edvfr}. This thread of research was motivated by the finding that direct estimation of the unobserved frequency count is numerically and statistically unstable, especially in high-diversity settings. The general technique involves finding a suitable functional transformation of the frequency counts, then performs prediction on the missing frequencies. Ideally the chosen functional transformation should be ``well-behaved'', probabilistically motivated, intuitive, and visually plausible. \cite{edvfr} argue that frequency ratios satisfy these criteria in microbial settings. The class of models that they fit to the frequency ratios is of the form 
\begin{equation}
 \frac{f_{j+1}}{f_j} = \frac{\beta_0 + \beta_1j + \ldots + \beta_pj^p}{1+\alpha_1j + \ldots+ \alpha_qj^q} + \varepsilon_j,\qquad j=1,\ldots,J. \label{ratiomodel}
\end{equation}
where $f_j$ represents the number of taxa in the sample observed $j$ times, $\varepsilon_j$ represents a error term arising through sampling variability, and $p, q \in \mathcal{N}$, $\beta_0, \ldots, \beta_p$, $\alpha_1, \ldots, \alpha_q \in \mathcal{R}$ are parameters to be estimated.  Willis \& Bunge fit their model using nonlinear least squares. Once parameter estimates are obtained, their prediction of the number of unobserved species is $\hat{f_0} = \frac{f_1}{\hat{\beta}_0},$ so that the estimate of the total richness is $\hat{C} = \hat{f_0} + \sum_{j \geq 1} f_j$. 

The technique of \cite{edvfr} is heavily dependent on the singleton count, $f_1$. Small changes in this value can drastically affect the richness estimate directly, via $f_1$ itself, or indirectly, via the parameter estimates. The latter case is especially problematic when the sample is dominated by rare species, in which case $\hat{\beta}_0$ is small, increasing $f_1/\hat{\beta}_0$ further. Furthermore, the standard error in the estimate may dramatically increase with $f_1$. Therefore, inaccuracy of the singleton count, while almost guaranteed in microbial applications, diminishes the utility of traditional richness estimation techniques. 

We propose to predict the singleton and the unobserved frequency count based on the remaining frequency counts. Our model is then as in Equation~(\ref{ratiomodel}), and the parameters are estimated via heteroskedastic nonlinear least squares, with the modification that now $j=2,\ldots,J$. We then predict the singleton count by substitution: $$\hat{f}_1 = \frac{f_2}{\frac{\sum_{i=0}^p \hat{\beta}_i}{1+\sum_{k=1}^q \hat{\alpha}_k}},$$ and the unobserved diversity by $\hat{f}_0 = \hat{f}_1/\hat{\beta}_0$. The resulting richness estimate is $\hat{C}=\hat{f}_0+\hat{f}_1 + \sum_{j\geq 2} f_j$. The standard error in this estimate is derived using the delta method in conjunction with a multinomial model for the frequency counts (see Supplementary Material for details).

The issue of appropriate model complexity is standard in statistics: an underparametrized model may only poorly fit the data while an overparametrized model does not permit prediction. Fortunately, the problem of diversity estimation admits a natural criterion for model selection: discard all models which imply negative unobserved diversity. In practice, we find that small models (eg., $p=1,q=0$) correspond to negative unobserved diversity estimates in the case of high diversity datasets (i.e., $f_2/f_1$ and $f_3/f_2$ are small, eg., less than 0.3). However, incrementally increasing $p$ and $q$ permits more flexibility in the model, which usually results in a positive estimate. Among all models that permit positive estimates, we choose the most parsimonious model. This choice is motivated by the visual finding that the smallest model generally appears to fit the frequency ratios without permitting arbitrary turning points, which often arise with highly parametrized models. 

\section*{Results \& Discussion}
\begin{table}[!t]
\caption{Comparison of the method breakaway\_nof1 with other species richness estimators: breakaway, CatchAll and Chao1. 20\% trimmed root-MSE is shown. True $C=5000$, and the frequency count distribution is negative binomial with probability $p=0.99$, size $n=500$ and density ${x+n-1 \choose x}p^n(1-p)^x$. The observed singleton count was increased by the percentage shown to mimic both false (chimeric) diversity and aggressive singleton filtering. Results based on 10,000 iterations. \label{nof1table}}
\begin{tabular}{lrrrr}\hline
Chimera/filtering rate & \url{breakaway_nof1} & breakaway & CatchAll & Chao1 \\ \hline
  100\% & 83.41 & 210.95 & 169.61 & 263.53 \\
  0\% & 83.39 & 5.00 & 4.49 & 5.98\\
  -80\% & 83.70 & 158.05 & 136.40 & 162.90 \\ \hline
\end{tabular}
\end{table}


We compare the species richness estimates obtained from \url{breakaway_nof1} to three species richness estimators in Table \ref{nof1table}: breakaway \citep{edvfr}, CatchAll \citep{catchall}, and Chao1 \citep{chao1}. To showcase the performance of \url{breakaway_nof1} in the presence of measurement error in the singletons, frequency count tables were constructed by simulating from a negative binomial distribution and zero-truncating the frequencies to reflect the unobserved OTUs. Three scenarios were considered: false diversity (singleton count altered to 200\% of its simulated value), correct singleton value (no alteration), and aggressive filtering (singleton count altered to 20\% of its simulated value). In the first and third instances, \url{breakaway_nof1} vastly outperforms its competitors with respect to 20\% trimmed root-MSE. However, \url{breakaway_nof1} is outperformed by other estimators when the correct singleton count is available, because in this case it ignores informative data. Note that classical MSE is inappropriate here because richness estimates display heavy right-skew.

breakaway\_nof1 may produce larger standard errors than other richness estimators, because if the singleton count can be accurately determined then \url{breakaway_nof1} ignores relevant data. In this way, Table

Runtime statistics, confirmation of standard errors under simulation, convergence properties, sample-based ``rarefaction'' convergence, and analysis of two soil microbiome datasets are available as Supplementary Material.
 
\section*{Conclusion}
The key contribution of \url{breakaway_nof1} is an easily accessible exploratory tool for investigating the robustness of species richness quantification to the extent of quality control employed in pre-processing community sequence data. As debate continues regarding the appropriate level of singleton filtering, this tool provides a statistically-motivated method for confirming the robustness of richness estimates.

\section*{Acknowledgement}
The author would like to thank John Bunge for many discussions that substantially improved this manuscript. The motivating problem was contributed by the attendees of the Marine Biological Laboratory's STAMPS workshop in 2013 and 2014 and the author is very grateful to those involved. 

\section*{Supplementary Material}

\subsection*{Standard error calculation}
I now outline the details of the standard error calculation pertaining to the species richness estimator \url{breakaway_nof1}.

Consider community composition data pertaining to $c$ observed taxa: random variables $X_1,\ldots,X_c$ where $\{X_i=j\}$ denotes the event that the $i$th taxon was observed $j$ times in the sample. We wish to estimate the {\it total diversity}: the number of taxa that exist in the population from which the sample was taken. The total diversity can be partitioned into the observed diversity (taxa present in the sample) and the unobserved diversity (taxa not present in the sample). Let $C$ denote the true number of distinct taxonomic classes in the population: we condition on the event that $\{X_{c+1}=0\} \cap \ldots \cap \{X_C=0\}$.

This data stucture can be reduced to the {\it frequency counts} $f_1,f_2,\ldots$, where $f_j = \#\{X_i=j\}, j=1,2,\ldots$. If we let $f_0$ denote the (unknown) number of species that were unobserved, then $C=f_0+f_1 + f_2 + \ldots$, and since $f_1,f_2,\ldots$ are known, estimating $C$ is equivalent to predicting $f_0$.

Now suppose that $f_1$ is corrupted: it is unclear whether this realization contains any information about the true number of taxa observed only once. We hence disregard its realization for use in estimation of $C$. However, there exists some true number of singletons in the population, and hence $f_1$ is now an unknown quantity to be predicted.

We consider the model for the {\it frequency ratios} $f_{j+1}/f_j$:
\begin{equation}
 \frac{f_{j+1}}{f_j} = \frac{\beta_0 + \beta_1j + \ldots + \beta_pj^p}{1+\alpha_1j + \ldots+ \alpha_qj^q} + \varepsilon_j,\qquad j=2,\ldots,J. \label{ratiomodel}
\end{equation}
where $\varepsilon_j$ represents a error term arising through sampling variability, and $p, q \in \mathbbm{N}$, $\beta_0, \ldots, \beta_p, \alpha_1, \ldots, \alpha_q \in \mathbbm{R}$ are parameters to be estimated. Our goal in this manuscript is to derive an estimate of $C$ and its standard error. We defer discussion of parameter estimation since the algorithm is identical to that described in detail in Willis \& Bunge (2015).

Suppose we estimate $p, q, \beta_0, \ldots, \beta_p, \alpha_1, \ldots, \alpha_q$ by $\hat{p}, \hat{q}, \hat{\beta}_0, \ldots, \hat{\beta}_{\hat{p}}, \hat{\alpha}_1, \ldots, \hat{\alpha}_{\hat{q}}$, and consider the prediction for $f_1$: $$\hat{f}_1 = \frac{f_2}{\frac{\sum_{i=0}^{\hat{p}} \hat{\beta}_i}{1+\sum_{k=1}^{\hat{q}} \hat{\alpha}_k}},$$ and the unobserved diversity prediction  $\hat{f}_0 = \hat{f}_1/\hat{\beta}_0$. The resulting richness estimate is $\hat{C}=\hat{f}_0+\hat{f}_1 + \sum_{j\geq 2} f_j$. We wish to find an expression for the variance in $\hat{C}$.

Let $n=f_2+f_3 + \ldots$. We write
\begin{align}
\var(\hat{C}) &= \var(\hat{f}_0+\hat{f}_1+n) \notag \\
&= \var\left(
\left( \begin{array}{ccc} 1 & 1 & 1 \end{array} \right)
\left( \begin{array}{c} \hat{f}_0 \\ \hat{f}_1 \\ n \end{array} \right)
\right) \notag \\
&= \left( \begin{array}{ccc} 1 & 1 & 1 \end{array} \right) \times \cov
\left( \begin{array}{c} \hat{f}_0 \\ \hat{f}_1 \\ n \end{array} \right) \times 
\left( \begin{array}{c} 1 \\ 1\\ 1\end{array} \right) \notag \\
&= \left( \begin{array}{ccc} 1 & 1 & 1 \end{array} \right) \times 
\left( \begin{array}{ccc} \var(\hat{f}_0) & \cov(\hat{f}_0,\hat{f}_1) & \cov(\hat{f}_0,n) \\ \cov(\hat{f}_0,\hat{f}_1)  &\var(\hat{f}_1) & \cov(\hat{f}_1,n) \\ \cov(\hat{f}_0,n) &\cov(\hat{f}_1,n) & \var(n) \end{array} \right) \times 
\left( \begin{array}{c} 1 \\ 1\\ 1\end{array} \right)
. \notag 
\end{align}
It remains to find expressions for each of these components. 

We consider a multinomial model for $f_0,f_1,\ldots$ with $C$ trials and event probabilities $p_0,p_1,\ldots$. Note that this does not contradict the model for the frequencies, which merely places additional structure on the event probabilities. Under this multinomial model, we have the following moments:
\begin{align}
\ev(f_i) &= Cp_i \notag \\
\var(f_i) &= Cp_i(1-p_i) \notag \\
\cov(f_i,f_j) &= -Cp_ip_j, \qquad i\neq j. \notag
\end{align}
We begin with the upper $2 \times 2$ components. Define $\hat{b}:= \frac{\sum_{i=0}^{\hat{p}} \hat{\beta}_i}{1+\sum_{k=1}^{\hat{q}} \hat{\alpha}_k}$, such that $\hat{f}_1 = \frac{f_2}{\hat{b}}.$ Then
\begin{align}
\cov(\hat{f}_0,\hat{f}_1) &= \cov\left(\frac{\hat{f}_1}{\hat{\beta}_0},\hat{f}_1 \right)  \notag  \\
&= \cov\left(\frac{f_2}{\hat{\beta}_0\hat{b}}, \frac{f_2}{\hat{b}} \right)  \notag  \\
&= \cov(g(f_2,\hat{\beta}_0,\hat{b})), \notag
\end{align}
where $g: \mathbbm{R}^3 \rightarrow \mathbbm{R}^2$, $g(f_2,\hat{\beta}_0,\hat{b}) := \left(\frac{f_2}{\hat{\beta}_0\hat{b}}, \frac{f_2}{\hat{b}} \right)$. Consider that 
$$\nabla g= \left(\begin{array}{cc} \frac{1}{\hat{\beta}_0\hat{b}} & \frac{1}{\hat{b}} \\ -\frac{f_2}{\hat{\beta}_0^2 \hat{b}} & 0 \\ -\frac{f_2}{\hat{\beta}_0\hat{b}^2} & -\frac{f_2}{\hat{b}^2}\end{array} \right).$$
We use a first order delta method approximation
$$\cov(g(f_2,\hat{\beta}_0,\hat{b})) \approx \nabla g^T \big|_{(\ev f_2, \ev \hat{\beta}_0,\ev \hat{b})} \times \cov(f_2,\hat{\beta}_0,\hat{b})\times  \nabla g\big|_{(\ev f_2, \ev \hat{\beta}_0,\ev \hat{b})}, $$ noting that $\ev f_2 = Cp_2$ under the multinomial model. Assume that $\ev \hat{\beta}_0 = \beta_0$, and $\ev \hat{b}=b$ for $b:=\frac{\sum_{i=0}^{p} \beta_i}{1+\sum_{k=1}^{q} \alpha_k}$. Then
\begin{align}
\cov(\hat{f}_0,\hat{f}_1) &\approx 
\left( \begin{array}{ccc} \frac{1}{\beta_0b}  & -\frac{Cp_2}{\beta_0^2 b} & -\frac{Cp_2}{\beta_0b^2} \\ \frac{1}{b} &0& -\frac{Cp_2}{b^2} \end{array} \right)
 \times \cov(f_2,\hat{\beta}_0,\hat{b}) \times
\left(\begin{array}{cc} \frac{1}{\beta_0b} & \frac{1}{b} \\ -\frac{Cp_2}{\beta_0^2 b} & 0 \\ -\frac{Cp_2}{\beta_0b^2} & -\frac{Cp_2}{b^2}\end{array} \right) \notag \\
&\approx 
\left( \begin{array}{ccc} \frac{1}{\beta_0b}  & -\frac{Cp_2}{\beta_0^2 b} & -\frac{Cp_2}{\beta_0b^2} \\ \frac{1}{b} &0& -\frac{Cp_2}{b^2} \end{array} \right) \times  \left( \begin{array}{ccc} Cp_2(1-p_2) & 0 & 0 \\ 0 &\var(\hat{\beta}_0) & \cov(\hat{b},\hat{\beta}_0) \\ 0 & \cov(\hat{b},\hat{\beta}_0) &\var(\hat{b}) \end{array} \right)\notag \\
 &\qquad  \times
\left(\begin{array}{cc} \frac{1}{\beta_0b} & \frac{1}{b} \\ -\frac{Cp_2}{\beta_0^2 b} & 0 \\ -\frac{Cp_2}{\beta_0b^2} & -\frac{Cp_2}{b^2}\end{array} \right) \notag \\
&=
\left( \begin{array}{ccc} 
\frac{Cp_2(1-p_2)}{\beta_0b} 
& -\frac{Cp_2}{\beta_0^2b}\var(\hat{\beta_0})-\frac{Cp_2}{\beta_0b^2}\cov(\hat{b},\hat{\beta}_0)
& -\frac{Cp_2}{\beta_0^2b}\cov(\hat{b},\hat{\beta_0})-\frac{Cp_2}{\beta_0b^2}\var(\hat{b}) \\
\frac{Cp_2(1-p_2)}{b} 
& -\frac{Cp_2}{b^2} \cov(\hat{b},\hat{\beta}_0)
& -\frac{Cp_2}{b^2}\var(\hat{b})
\end{array} \right)   \notag \\ &\qquad \times
\left(\begin{array}{cc} \frac{1}{\beta_0b} & \frac{1}{b} \\ -\frac{Cp_2}{\beta_0^2 b} & 0 \\ -\frac{Cp_2}{\beta_0b^2} & -\frac{Cp_2}{b^2}\end{array} \right), \notag 
\end{align}
which simplifies to
\begin{align}
\var(\hat{f}_0) &=
\frac{Cp_2(1-p_2)}{\beta^2_0b^2} +\frac{C^2p_2^2}{\beta_0^4b^2}\var(\hat{\beta_0})+2\frac{C^2p_2^2}{\beta_0^3b^3}\cov(\hat{b},\hat{\beta}_0) + \frac{C^2p_2^2}{\beta_0^2b^4} \var(\hat{b})   \notag \\
\cov(\hat{f}_0,\hat{f}_1) &= \frac{Cp_2(1-p_2)}{\beta_0 b^2} + \frac{C^2p_2^2}{\beta_0^2 b^3} \cov(\hat{b},\hat{\beta}_0) + \frac{C^2p_2^2}{\beta_0b^4}\var(\hat{b}) \notag \\
\var(\hat{f}_1) &= \frac{Cp_2(1-p_2)}{b^2} + \frac{C^2p_2^2}{b^4}\var(\hat{b}). \notag
\end{align}
Since all of the RHS terms above are unknown, we substitute in our empirical estimates to find our final estimates. In doing so, we assume that the estimates are ``close'' to their true values. 
\begin{align}
\hat{\var}(\hat{f}_0) &=
\frac{f_2(\hat{C}-f_2)}{\hat{C}\beta^2_0b^2} +\frac{f_2^2}{\beta_0^4b^2}\hat{\var}(\hat{\beta_0})+2\frac{f_2^2}{\beta_0^3b^3}\hat{\cov}(\hat{b},\hat{\beta}_0) + \frac{f_2^2}{\beta_0^2b^4} \hat{\var}(\hat{b})   \notag \\
\hat{\cov}(\hat{f}_0,\hat{f}_1) &= \frac{f_2(\hat{C}-f_2)}{\hat{C}\beta_0 b^2} + \frac{f_2^2}{\beta_0^2 b^3} \hat{\cov}(\hat{b},\hat{\beta}_0) + \frac{f_2^2}{\beta_0b^4}\hat{\var}(\hat{b}) \notag \\
\hat{\var}(\hat{f}_1) &= \frac{f_2(\hat{C}-f_2)}{\hat{C}b^2} + \frac{f_2^2}{b^4}\hat{\var}(\hat{b}). \notag
\end{align}
where we find the regression coefficient empirical variances from the Hessian matrix obtained in their estimation. 

We turn our attention to the remaining three terms. Under the multinomial model we have
\begin{align}
\var(n) &= C(1-p_0-p_1)(p_0+p_1) \notag \\
\cov(\hat{f}_0,n) &= -Cp_0(1-p_0-p_1) \notag \\
\cov(\hat{f}_1,n) &= -Cp_1(1-p_0-p_1) \notag
\end{align}
and hence we use the plug-in estimates:
\begin{align}
\hat{\var}(n) &= \frac{n(\hat{f}_0+\hat{f}_1)}{\hat{C}} \notag \\
\hat{\cov}(\hat{f}_0,n) &= -\frac{\hat{f}_0n}{\hat{C}} \notag \\
\hat{\cov}(\hat{f}_1,n) &= -\frac{\hat{f}_1n}{\hat{C}}. \notag
\end{align}
Combining all of these terms together gives the required estimate of the variance (and therefore the standard error) of $\hat{C}$.

\subsection*{Performance under negative binomial models}

In Table \ref{errortab} I compare mean, median and 20\% trimmed mean square error of \url{breakaway_nof1} to breakaway, CatchAll, and Chao1. As is to be expected, we observe that \url{breakaway_nof1}'s best performance is when the chimeric rate (inflation of the singleton value as a percentage) differs greatly from zero. In all cases of chimeric rate greater than 40\%, \url{breakaway_nof1} outperforms its competitors.  Differences are marginal for 40\% singleton deflation. 

In every simulation, \url{breakaway_nof1}'s mean and trimmed mean square errors are greater than the median, suggesting that occasionally \url{breakaway_nof1} will produce large estimates for all data structures. See Section \ref{sec} for more discussion on the choice of comparative statistics. 

\begin{table}[h]
\caption{ 20\% trimmed mean, mean and median square error in estimating $\hat{C}$ using breakaway\_nof1 when the true frequency count distribution is negative binomially distributed with probability parameter $p$, size parameter $n$ and density  ${x+n-1 \choose x} p  /  n(1-p)^x$ with singleton inflation. Results are based on 500 replications. \label{errortab}}
\begin{tabular}{ r | r |r | r | r | r  }    
  ($C, n, p$, chimeric rate) & Statistic & \url{breakaway_nof1} & breakaway& CatchAll & Chao1 \\ \hline
$( 5000 , 500 , 0.99 , -80  ) $ & t-mean & 80.3 & 158.5 & 136.6 & 163\\
$( 5000 , 500 , 0.99 , -80  ) $ & mean & 99.5 & 274.8 & 137.1 & 163.5\\
$( 5000 , 500 , 0.99 , -80  ) $ & median & 75.6 & 158.3 & 135.9 & 162.6\\
$( 5000 , 500 , 0.99 , -40  ) $ & t-mean & 87.2 & 84.4 & 68.2 & 87.2\\
$( 5000 , 500 , 0.99 , -40  ) $ & mean & 101.8 & 84.7 & 68.8 & 87.5\\
$( 5000 , 500 , 0.99 , -40  ) $ & median & 80.6 & 84.4 & 68.7 & 87\\
$( 5000 , 500 , 0.99 , 0  ) $ & t-mean & 78.5 & 5.2 & 4.5 & 6.1\\
$( 5000 , 500 , 0.99 , 0  ) $ & mean & 94.8 & 6.8 & 6 & 8\\
$( 5000 , 500 , 0.99 , 0  ) $ & median & 73.2 & 4.6 & 4 & 5.5\\
$( 5000 , 500 , 0.99 , 50  ) $ & t-mean & 83.3 & 104.9 & 85.4 & 124.4\\
$( 5000 , 500 , 0.99 , 50  ) $ & mean & 103.7 & 105.7 & 85.6 & 126.2\\
$( 5000 , 500 , 0.99 , 50  ) $ & median & 76.7 & 104.8 & 85.1 & 124.4\\
$( 5000 , 500 , 0.99 , 100  ) $ & t-mean & 78.9 & 209.3 & 168.6 & 260.7\\
$( 5000 , 500 , 0.99 , 100  ) $ & mean & 96.4 & 211.1 & 169.6 & 264.2\\
$( 5000 , 500 , 0.99 , 100  ) $ & median & 72.8 & 208.8 & 168.4 & 260\\
  \hline  
\end{tabular}
\end{table}

We see that \url{breakaway_nof1} can be uniformly outperformed when no singleton inflation occurs. This highlights that accurate determination of the singleton count is vastly preferable to post-hoc procedures for its imputation or estimation. In this case, \url{breakaway_nof1} ignores valid data, thus reducing the effective sample size available and increasing its variability relative to other estimators. This variability is showcased above in Table \ref{errortab} by the tenfold increase in MSE compared to the other estimators in the zero chimeric singleton case. 

It is important to clarify that the simulations described do not reflect draws from a 1-inflated negative binomial distribution. Such a model would reflect the sampling procedure of drawing from the negative binomial distribution with probability $p$ and observing a chimera with probability $1-p$. The decision to instead simulate from a negative binomial distribution, and then, conditional on the observed $f_1$ (singleton count), add an additional (chimeric rate $\times f_1$) singletons to the sample reflects a data generating process whereby the more (true) rare species in the sample, the greater the risk of chimeras forming. The author believes that chimeric rates are conditional on the structure of the microbial sample and no universal chimeric rate exists, and as a result that the conditional singleton inflation model better reflects the data generating process of high throughput sequencing.

These size and probability parameter values were chosen based on the author's experience with fitting negative binomial distributions to {\it microbial} frequency count data. Microbial datasets contain greater numbers of singletons compared to macroecology abundance datasets, and thus their fitted probability parameters tend to be very close to 1 (the boundary case). Results for (size, probability) of $(100, 0.95)$ were extremely similar and can be obtained from the author upon request; alternatively, the simulation \url{.R} file is also available to the reader. A greater range of parameters were used to assess standard error reporting in Table \ref{stderrortab}. Additionally, it is typical for most microbial richness estimates to be in the range 2000 to 8000, hence the chosen true $C$ of 5000. Table \ref{stderrortab} also investigates robustness to $C$.

\subsection*{Standard error confirmation} \label{sec}

We observe in Table \ref{stderrortab} that standard errors are approximately correct to within 33\%, as sampled over a broad range of possible negative binomial models that reflect the high diversity (i.e. $p$ close to 1) seen in microbial studies. Interestingly, while standard errors for \url{breakaway_nof1} are large in general, they tend to slightly underreport their true variability for higher diversity parameter values (probability parameters closer to 1). On the other hand, in lower diversity instances the variability is overstated on average. This highlights the limitations of 1st order approximations to the variance functions of right-skewed estimators (see next paragraph). Note that introducing chimeras is not necessary in this simulation: \url{breakaway_nof1} (by nature) ignores the singleton count, and thus any differences in performance under different chimeric rates would be wholly attributed to sampling variability from drawing negative binomial samples. 

In conventional (i.e. non-boundary value) statistical inference problems, it is common to use the mean of the standard errors and compare it to the standard deviation of the estimates. However, this choice arises because most estimators follow an asymptotically normal distribution around the estimand. No such asymptotics apply in the species problem. Log-normal distributions are commonly used to model richness estimates, however this is merely heuristic. In order to reflect the heavy tails of richness estimates and present a more realistic picture of \url{breakaway_nof1}'s error distribution, I have evaluated it with respect to median (rather than mean) and its equivalent variability statistic, median absolute difference (rather than standard deviation). Table \ref{errortab} showcases the vast differences between the mean and median squared error in determining accuracy.

\begin{table}[h]
\begin{center}
\caption{\label{stderrortab}   Median estimated standard error and actual median absolute deviation in estimating $\hat{C}$ using breakaway\_nof1 when the true frequency count distribution is negative binomially distributed with probability parameter $p$, size parameter $n$ and density  ${x+n-1 \choose x} p  /  n(1-p)^x$. Results are based on 5,000 replications.}
\begin{tabular}{ r | r | r | r } \hline       
  ($C, n, p$) &  $\hat{\mbox{s.e.}}(\hat{C})$ & True s.e.($\hat{C}$)  & Error \\ \hline 
$( 8000 , 500 , 0.99  ) $ &102.69 & 125.06 & -17.89\% \\
$( 8000 , 100 , 0.95  ) $ &178.32 & 163.25 & 9.23\% \\
$( 5000 , 500 , 0.99  ) $ &63.74 & 82.22 & -22.48\% \\
$( 5000 , 100 , 0.95  ) $ &119.11 & 107.86 & 10.43\% \\
$( 3000 , 500 , 0.99  ) $ &38.42 & 58.06 &  -33.83\% \\
$( 3000 , 100 , 0.95  ) $ &77.30 & 72.83 & 6.14\% \\  \hline  
\end{tabular}
\end{center}
\end{table}

\subsection*{Data analysis}

We compare the species richness estimates obtained from \url{breakaway_nof1} on microbiome data in Table \ref{nof1table}. Two publicly available datasets were examined: an apple orchard soil microbiome \citep{apples} and a Hawaiian soil microbiome \citep{hawaii}. In each case we observe that the \url{breakaway_nof1} estimates were comparable with other procedures after accounting for the standard error in the estimates. However, in both cases the standard errors are very large, reflecting the fundamental difficulty in estimating the number of unobserved species with no knowledge of the singleton count. Note that for the Hawaiian soil dataset, even CatchAll (which notoriously underreports standard errors) cites a relatively large standard error. We note that the Chao1 estimates tend to be lower than the other estimates and with lower standard errors, which reflects that the Chao1 estimator reflects a lower bound (rather than an estimate) under the assumption of mixed Poisson frequency counts. 

Both datasets are available in the R package.

\begin{table}[h]
\caption{Comparison of the method breakaway\_nof1 with other species richness estimators: breakaway, CatchAll and Chao1. Standard errors are given in parentheses.\label{nof1table}}
\begin{tabular}{lllll} \hline
Dataset & \url{breakaway_nof1} & breakaway & CatchAll & Chao1 \\ \hline
Apple orchard soil & 1500 (1341) & 1552 (305) & 1477 (59) & 1241 (38) \\
Hawaiian soil & 3100 (2944)  & 5772 (4217) & 4887 (683) & 3182 (89)\\\hline
\end{tabular}
\end{table}


\subsection*{Runtime analysis}

In Table \ref{times} we observe that \url{breakaway_nof1} and breakaway have comparable runtimes of between 0.08 and 0.21 seconds. It is of interest to note that breakaway becomes slightly slower when the singleton rate increases. This is likely due to its procedure needing to fit increasingly higher order models in order to satisfy the boundary condition $\hat{f}_0>0$. CatchAll's runtimes are uniform across the various parameter values at approximately 0.44 seconds (two to four times slower than \url{breakaway_nof1} and breakaway). 

Runtimes are only given for the computationally intensive estimators. Note that Chao1's runtimes are essentially negligible since only a single division needs to be performed. 

These simulations were conducted on a 2.7 GHz Intel Core i5 2015 MacBook Pro; the user's experience with runtimes may differ with his/her computational resources. 

\begin{table}[h]
\caption{ \label{times} Run times for estimating $\hat{C}$ using breakaway\_nof1 compared to a number of additional competitors when the true frequency count distribution is NB with probability $p$, size $n$ and density  ${x+n-1 \choose x} p  /  n(1-p)^x$ with singleton inflation. Results are based on 500 replications.}
\begin{tabular}{ r | r |r | r | r   }    
  ($C, p, n$, chimeric rate) & Statistic & \url{breakaway_nof1} & breakaway& CatchAll  \\ \hline
$( 5000 , 500 , 0.99 , -80  ) $ & t-mean & 0.114 & 0.079 & 0.439\\
$( 5000 , 500 , 0.99 , -80  ) $ & mean & 0.118 & 0.098 & 0.44\\
$( 5000 , 500 , 0.99 , -80  ) $ & median & 0.114 & 0.07 & 0.439\\
$( 5000 , 500 , 0.99 , -40  ) $ & t-mean & 0.114 & 0.112 & 0.436\\
$( 5000 , 500 , 0.99 , -40  ) $ & mean & 0.122 & 0.123 & 0.437\\
$( 5000 , 500 , 0.99 , -40  ) $ & median & 0.112 & 0.109 & 0.437\\
$( 5000 , 500 , 0.99 , 0  ) $ & t-mean & 0.114 & 0.11 & 0.435\\
$( 5000 , 500 , 0.99 , 0  ) $ & mean & 0.119 & 0.115 & 0.435\\
$( 5000 , 500 , 0.99 , 0  ) $ & median & 0.113 & 0.109 & 0.436\\
$( 5000 , 500 , 0.99 , 50  ) $ & t-mean & 0.115 & 0.147 & 0.439\\
$( 5000 , 500 , 0.99 , 50  ) $ & mean & 0.12 & 0.153 & 0.442\\
$( 5000 , 500 , 0.99 , 50  ) $ & median & 0.115 & 0.145 & 0.438\\
$( 5000 , 500 , 0.99 , 100  ) $ & t-mean & 0.114 & 0.205 & 0.439\\
$( 5000 , 500 , 0.99 , 100  ) $ & mean & 0.127 & 0.21 & 0.441\\
$( 5000 , 500 , 0.99 , 100  ) $ & median & 0.113 & 0.203 & 0.437\\ \hline  
\end{tabular}
\end{table}

\subsection*{Convergence under simulation}

\begin{center}
\begin{figure}[h]
\includegraphics[scale=0.6,page=1]{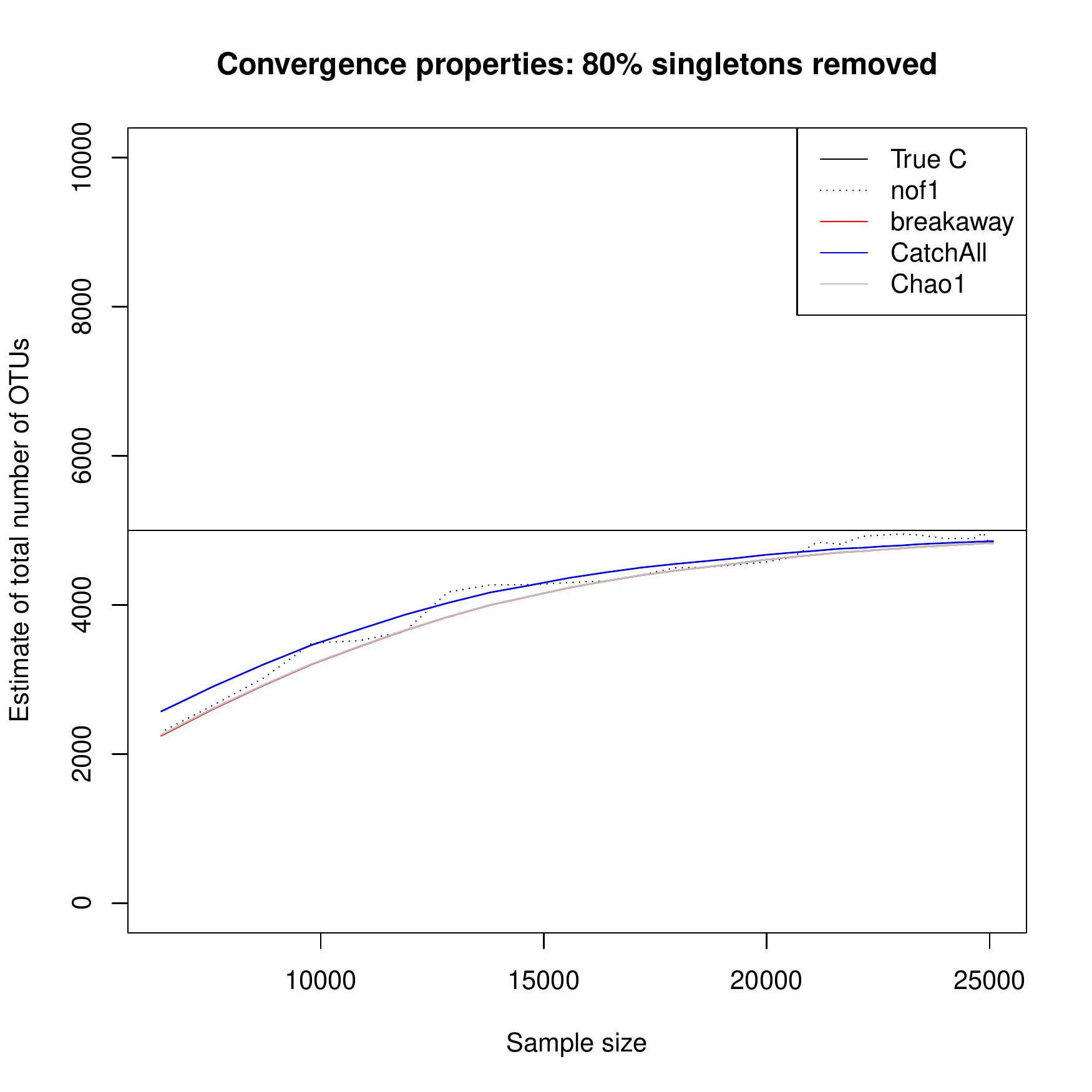}
\caption{Convergence of 4 species richness estimators when the true frequency count distribution is negative binomial with probability $0.99$, size $5000$ and $C=5000$ with 80\% singleton deflation}
\label{rare0}  
\end{figure}
\end{center}

\begin{center}
\begin{figure}[h]
\includegraphics[scale=0.6,page=2]{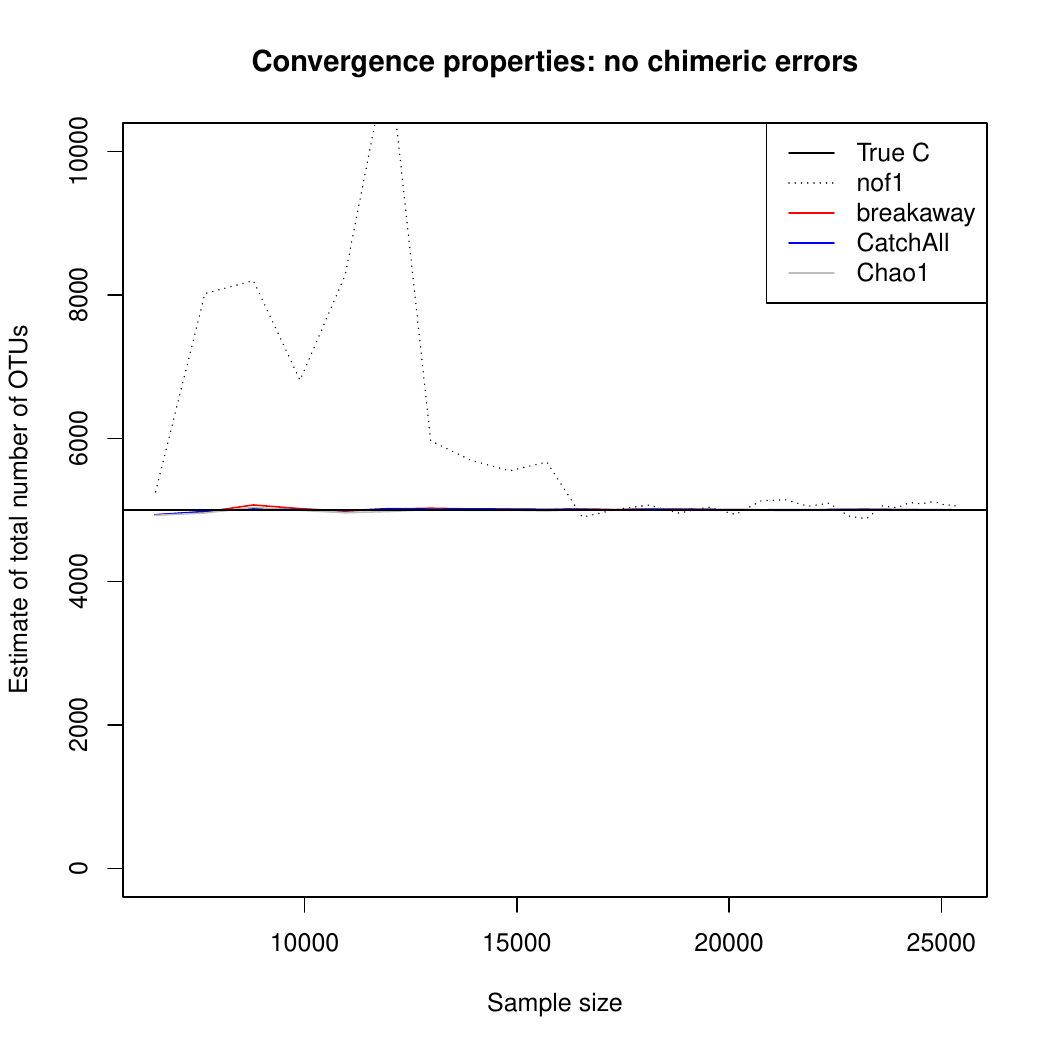}
\caption{Convergence of 4 species richness estimators when the true frequency count distribution is negative binomial with probability $0.99$, size $5000$ and $C=5000$ with no singleton deflation. }
\label{rare1}  
\end{figure}
\end{center}

\begin{center}
\begin{figure}[h]
\includegraphics[scale=0.6,page=3]{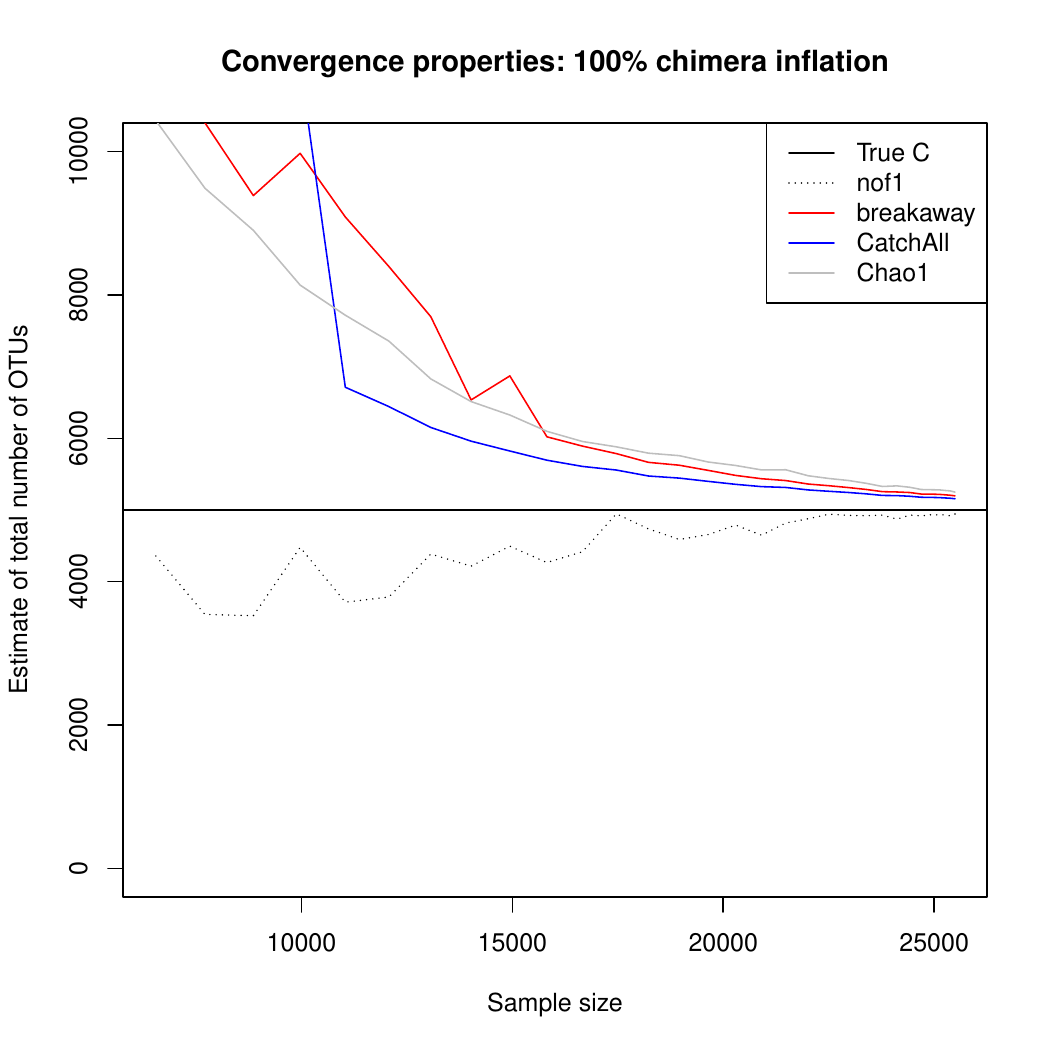}
\caption{Convergence of 4 species richness estimators when the true frequency count distribution is negative binomial with probability $0.99$, size $5000$ and $C=5000$ with 100\% singleton deflation. }
\label{rare2}  
\end{figure}
\end{center}

In Figures \ref{rare0}, \ref{rare1} and \ref{rare2}, we observe the convergence behavior of the 4 different richness estimates under different levels of singleton deflation and inflation. In the deflation case (Figure\ref{rare0}), \url{breakaway_nof1} converges at approximately the same rate as the other estimators, and from below (convergence from below is common in the species problem, and gives rise to the common misperception that the true richness is a function of the sample size). In the case of no errors (Figure \ref{rare1}), \url{breakaway_nof1} is unstable for small sample sizes but converges towards the correct $C$ of 5000. In contrast, when chimera inflation is large (Figure \ref{rare2}), \url{breakaway_nof1} certainly converges first, but does so from below, while other estimators do so from above. Overall it appears that in the presence of singleton uncertainty, \url{breakaway_nof1} performs no worse with respect to convergence, and substantially better with respect to error (Table \ref{errortab}).

\subsection*{Convergence: sample-based rarefaction} 

\begin{center}
\begin{figure}[h]
\includegraphics[scale=0.6]{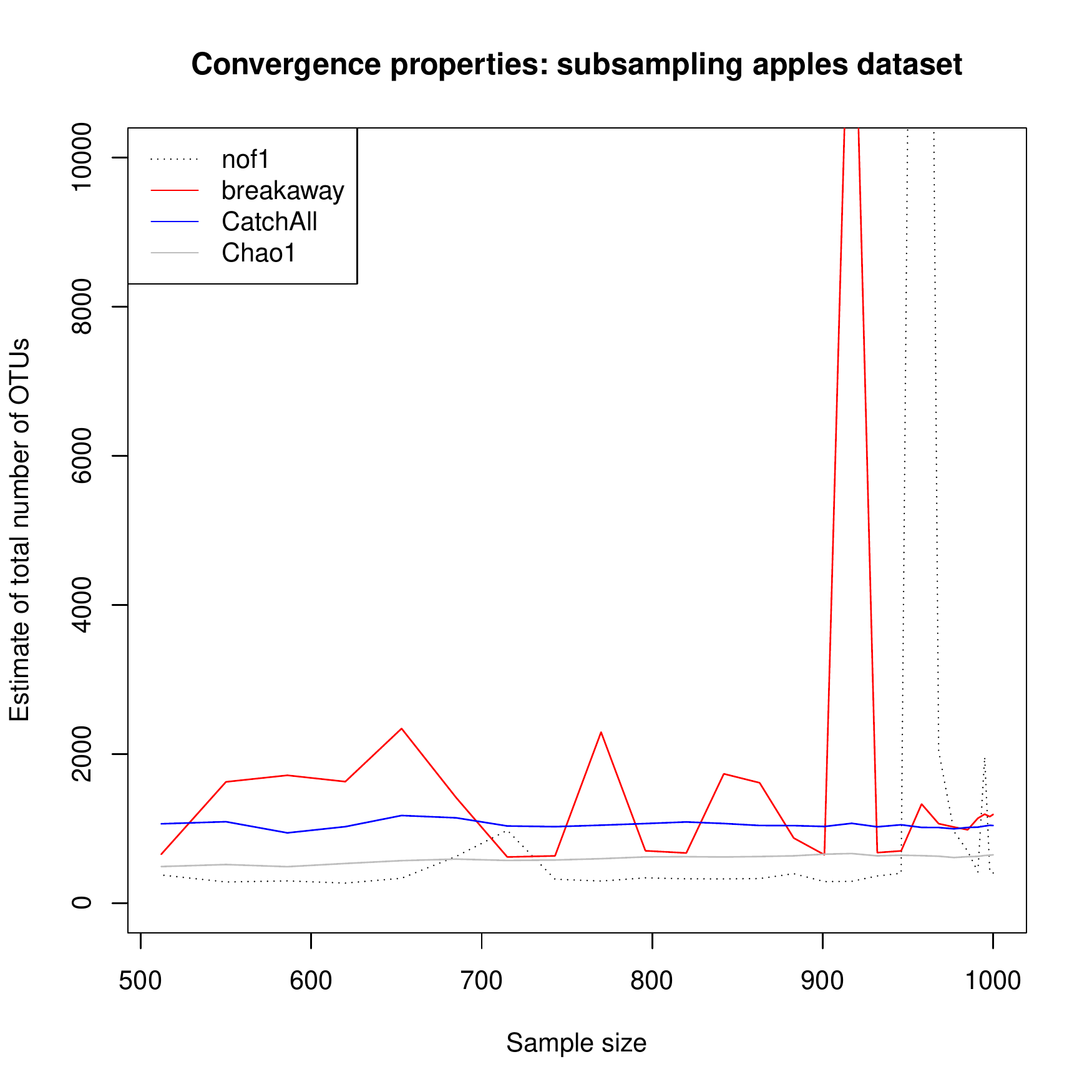}
\includegraphics[scale=0.6]{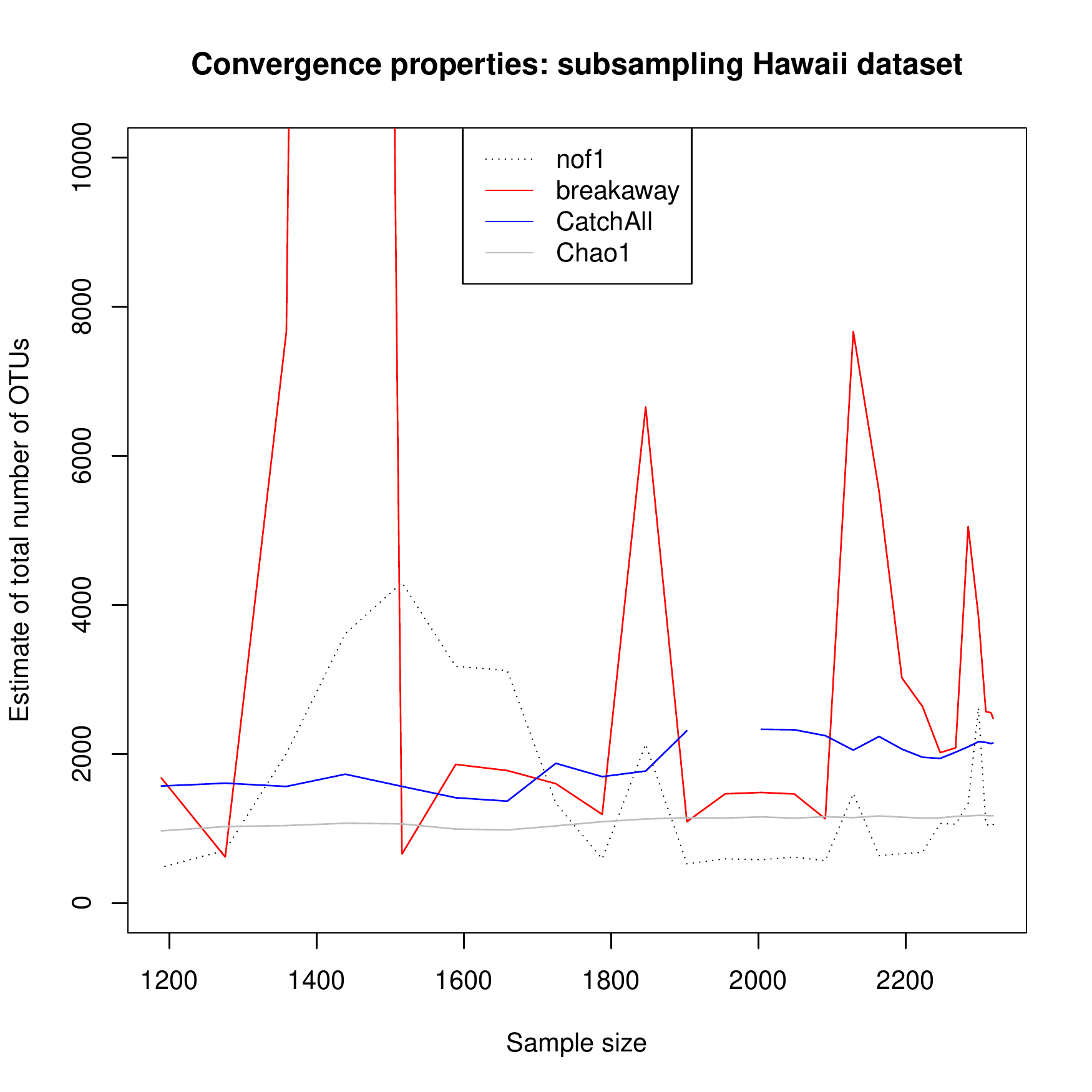}
\caption{Convergence under subsampling of 4 species richness estimators for the Apples and Hawaii datasets.}
\label{apples}  
\end{figure}
\end{center}

Sample-based rarefaction curves for the Apples and Hawaii datasets are visible in Figure \ref{apples}. Unsurprisingly, substantially more volatility is observed in these curves than in the negative binomially simulated curves. It appears that \url{breakaway_nof1} is more stable across subsampling compared to breakaway, but less so than CatchAll and Chao1 (again, unsurprisingly). Since rarefaction-based inference is statistically inadmissible (the average error of a rarefied estimator can always be beaten by a non-rarefied estimator), I am unwilling to draw any conclusions about the true richnesses of the populations based on these plots. I hope the reader is similarly dissuaded by the outcry of statisticians against rarefaction-based inference. For an accessible discussion, see \cite{wnwn}.

\bibliography{EDvFR}

\end{document}